\documentclass[aps,prb,twocolumn,superscriptaddress,preprintnumbers]{revtex4-2}

\usepackage[utf8]{inputenc}
\usepackage[T1]{fontenc}
\usepackage{amsmath}
\usepackage{amsfonts}
\usepackage{amssymb}
\usepackage{graphicx}
\usepackage{bbold}
\usepackage{physics}

\usepackage{xcolor}
\usepackage{hyperref}
\usepackage{braket}
\usepackage{subcaption}

\begin{document}


\title{Effects of defects in the XY chain with frustrated boundary conditions}


\author{Gianpaolo Torre}
\affiliation{Department of Physics, Faculty of Science, University of Zagreb, Bijeni\v{c}ka cesta 32, 10000 Zagreb, Croatia.}
\author{Vanja Mari\'{c}}
\affiliation{Division of Theoretical Physics, Ru{\dj}er Bo\v{s}kovi\'{c} Institute, Bijeni\v{c}ka cesta 54, 10000 Zagreb, Croatia}
\affiliation{SISSA and INFN, via Bonomea 265, 34136 Trieste, Italy}
\author{Fabio Franchini}
\affiliation{Division of Theoretical Physics, Ru{\dj}er Bo\v{s}kovi\'{c} Institute, Bijeni\v{c}ka cesta 54, 10000 Zagreb, Croatia}
\author{Salvatore Marco Giampaolo}
\affiliation{Division of Theoretical Physics, Ru{\dj}er Bo\v{s}kovi\'{c} Institute, Bijeni\v{c}ka cesta 54, 10000 Zagreb, Croatia}


\date{\today}

\begin{abstract}
It has been recently proven that new types of bulk, local order can ensue due to frustrated boundary condition, that is, periodic boundary conditions with an odd number of lattice sites and anti-ferromagnetic interactions. For the quantum XY chain in zero external fields, the usual antiferromagnetic order has been found to be replaced either by a mesoscopic ferromagnet or by an incommensurate AFM order. In this work we examine the resilience of these new types of orders against a defect that breaks the translational symmetry of the model. We find that, while a ferromagnetic defect restores the traditional, staggered order, an AFM one stabilizes the incommensurate order. The robustness of the frustrated order to certain kinds of defects paves the way for its experimental observability.
\end{abstract}

\preprint{RBI-ThPhys-2020-21}


\maketitle

\section{Introduction}

Landau's paradigm constitutes a cornerstone for the understanding of phases of many-body systems~\cite{Landau1978}. 
It classifies different phases through the analysis of local order parameters that, assuming a non-zero value, signal the rise up of specific orders.
In classical settings, this paradigm allows the complete classification of the different phases.
But, although Landau's theory remains an indispensable tool in the quantum regime, it does not allow us to grasp all the richness and variety of quantum many-body physics, whose nature, being non-local, does not necessarily fit into the paradigm. 
Emblematic examples of such violation are represented by the topological and nematic ordered phases~\cite{Stone1992, Nayak2008, Fradkin2013, Bernevig2013, Witten2016, Zeng2019, Wen2004, Hasan2010, Giampaolo2015, Lacroix2011}, that, in disagreement with Landau's paradigm, are not characterized by local order parameters that violate some symmetries of the system.

While this is a well-known limit of Landau's theory in the quantum regime, in the last years other problems were brought to light.
Indeed, a clear statement of Landau's theory is that the thermodynamic properties of the different phases must be independent of the boundary conditions whose contribution is expected to be subdominant compared to the bulk interactions~\cite{Burkhardt1985, Cabrera1986, Cabrera1987}. 
This consideration, which corresponds to our classical intuition of complex systems, has been proven to be wrong in the quantum regime.
On one hand, it was shown that, in general, the knowledge of the system at finite size is not sufficient to determine its spectral gap properties in the thermodynamic limit~\cite{Cubitt2015, Bausch2018}. 
Furthermore, an explicit model has been constructed in which, tuning the interaction between the edges of an open chain, the system goes through a quantum phase transition~\cite{Campostrini2015}.
Moreover, recently, following the same line of research, it has been shown 
that frustrated boundary conditions (FBC), namely the case of periodic boundary conditions with an odd number of spins, associated to an antiferromagnetic short-range interaction destroys local order~\cite{Maric20, MaricToeplitz} and induces a first-order quantum phase transition (QPT) that is absent when other boundary conditions are considered~\cite{Maric20_2}.

Indeed, the assumption of the FBC, even in the classical regime, has a deep impact on the ground state (GS) properties of the system.
While in the case of ferromagnetic (FM) interaction all local terms in the Hamiltonian can be minimized simultaneously, in the presence of antiferromagnetic coupling (AFM), due to the oddness of chain sites, at least one bond needs to be aligned ferromagnetically. 
The effect FBC is then to force the competition between incompatible orderings, resulting in the rising of a frustration~\cite{Toulouse1977,Vannimenus1977,Wolf2003,Giampaolo2011,Giampaolo2015_2,Marzolino2013,Sadoc2007,Lacroix2011,Diep20013} of topological nature in the system.
As a consequence, in the classical case, such as the classical Ising chain with AFM interactions, FBC induce, starting from one of the two Neel states, $2N$ degenerate lower energy kink states, each one of them characterized by a different position of a magnetic defect, i.e. two spins parallelly oriented in a Neel state.
The effect of the quantum interaction is to lift this degeneracy, generating a Galilean band of gapless excitations in contact with the lower energy state(s)~\cite{Dong2016, Dong2017, Dong2018, Li2019}. 
It is possible to show that the system can be characterized by the existence, on top of the frustrated GS, of a delocalized excitation along the chain~\cite{Giampaolo2019, Maric20}, that destroys the AFM local order. 

Since quantum interactions tend to delocalize the magnetic defect in the whole system, it is natural to wonder what happens with the introduction of a localized defect in the interaction pattern that explicitly breaks the translational invariance.
In general, it is known that the presence of defects in a spin chain can induce a very rich phenomenology~\cite{Schuster2002} and can influence the system geometry~\cite{Apollaro2013}. 
In particular, in~\cite{Campostrini2015} it was shown that the case of FBC with perfect translational invariance is a first-order phase transition separating a {\it magnetic phase} when a defect favors a ferromagnetic order on a bond, from a {\it kink phase} with an AFM defect. The two phases were characterized by a difference scaling in the closing of the energy gap: Exponential in the magnetic phase and algebraic for the kink one.

In the present work, extending the analysis in~\cite{Maric20, Maric20_2}, we investigate the fate of the novel local orders in the two frustrated phases of an AFM XY chain with FBC in the presence of a localized defect. 
Under usual conditions, one does not expect that such defect can affect the system beyond some finite distance around it. Even more, since the ground state with FBC is interpreted as a single excitation state, the effect of a defect could be to localize this excitation, thus restoring the traditional order, except for an exponentially limited area whose relevance, in the thermodynamic limit, becomes negligible. 
These considerations are probably one of the reason for which the aforementioned orders emerging with FBC have been overlooked for too long: they have been expected to be too weak against defect and thus impossible to detect experimentally.
We will show that this picture is correct only when a ferromagnetic type defect (FTD), i.e. a defect that reduces the relative weight of the dominant AFM term, is considered. 
On the contrary, when we take into account an antiferromagnetic type defect (AFTD), i.e. a defect that locally increases the dominant AFM term, an incommensurate AFM order is induced in the system. 
This incommensurate AFM order holds a magnetic pattern very close to the one in~\cite{Maric20_2} but, differently from it, is associated with a two-fold degenerate ground-state and not a four-fold one.
Thus, while the mesoscopic ferromagnetic order described in~\cite{Maric20} does not seem to survive in presence of any defect, the incommensurate AFM order is found to be resilient also to the presence of a second defect, indicating that it can be observed under relatively general conditions with FBC. 
The emergence of two different orders (i.e. the standard AFM and the incommensurate staggered ones) signals the existence of a quantum critical point (QPT) separating them. Contrary to \cite{Campostrini2015}, our bulk control parameter $\phi$ can cross a QPT that does not destroy the effects of frustration. In fact, in \cite{Campostrini2015}, the effect of the defect is considered only within a given phase, while, by varying $\phi$, we can move from a region with mesoscopic ferromagnetic order to one with incommensurate AFM order. Thus, the QPT we observe with the defect borrows its phenomenology both from \cite{Campostrini2015} and \cite{Maric20_2}. Most importantly, its existence relies on the loop geometry of the chain:
If we open the lattice, regardless of the nature of the defect, an almost perfect standard staggerization is restored in the bulk.
Hence, also in this case, the assumption of FBC push the system outside the range of validity of Landau's theory.

The paper is organized as follows. In Section~\ref{TheModel} we introduce the model under study and briefly review its properties in the absence of defects. 
In Section~\ref{Methods} we describe the analytical and numerical techniques we use to analyze the effects of adding the defect, which requires particular care, due to the closing of the gap with the lowest energy band in the thermodynamic limit.
In Section~\ref{Results} we show and discuss the results for various types of perturbations. Conclusions and outlook are collected in Section~\ref{Conclusions}. 

\section{The model}\label{TheModel}

All along within this paper, we focus on the XY chain at zero fields with FBC and a local defect that, without any loss of generality, we set between the first and the last spin of the chain. 
Such a system is described by the following Hamiltonian:
\begin{align}\label{HamDef}
	H & = \sum_{j=1}^{N-1} \cos\phi\;\sigma_{j}^x\sigma_{j+1}^x+\sum_{j=1}^{N-1} \sin\phi\;\sigma_j^y\sigma_{j+1}^y + \nonumber \\
	& +\cos(\phi+\delta_x)\;\sigma_N^x\sigma_1^x+\sin(\phi+\delta_y)\;\sigma_N^y\sigma_1^y,
\end{align}
where $\sigma_j^\alpha$, for $\alpha=x,y,z,$ are Pauli operators defined on the $j$-th spin and the FBC are achieved by imposing periodic boundary conditions $\sigma_{N+j}^\alpha\equiv\sigma_j^\alpha$ and an odd number $N$ of lattice sites. 
The parameter $\phi$ tunes the relative strength between the interactions along the $x$ and $y$ directions, while $\delta_x$ and $\delta_y$ govern the strength of the defect along the $x$ and $y$ axis respectively. 
The presence of the defect in the interaction pattern destroys the translational invariance of the model as well as all its mirror symmetries except the one respect to the $(N+1)/2$-th spin.

This is not the most general defect that we can consider. 
The reason behind our choice is that the Hamiltonian in eq.~\eqref{HamDef} still preserves the parity symmetries, $[H,\Pi^\alpha]=0$ with $\Pi^\alpha=\bigotimes^N_{j=1}\sigma_j^\alpha$, with respect to all the three spin directions, $\alpha=x,y,z$, as the unperturbed model.
This property is of particular relevance in our analysis, because it implies an exact degeneracy for the ground state already in a finite system.
Indeed, since $N$ is odd, parity operators anti-commute ($\{\Pi^\alpha,\Pi^\beta\}=2\delta_{\alpha\beta}$).
Hence, if the state $\ket{\varphi}$ is an eigenstate of both the Hamiltonian and one of the parity operators, say $\Pi^z$, the state $\Pi^x\ket{\varphi}$ is still an eigenstate of both $H$ and $\Pi^z$ but has the opposite $\Pi^z$ eigenvalues.
Hence we can conclude that each eigenstate of the Hamiltonian is (at least) two-fold degenerate even for finite size.
This degeneracy enables us to study the magnetization directly, even if a finite system, exploiting the trick introduced in~\cite{Maric20, Maric20_2}.

Before starting our analysis, let us briefly review here the main findings of the unperturbed model~\cite{Maric20, Maric20_2}, that corresponds to $\delta_x\!=\!\delta_{y}\!=\!0$ in eq.~\eqref{HamDef}. 
For $\phi$ in the region $(-3\pi/4,-\pi/4)$ the dominant term is the ferromagnetic interaction along the $y$ direction ($y$FM phase).
In the thermodynamic limit, the two-fold degenerate ground state manifold is separated from the rest of the spectrum by a finite energy gap and admits a ferromagnetic magnetization along y $m_y=\braket{\sigma_j^y}$.
This picture is completely equivalent to the one that can be found taking into account open boundary conditions or an even number of spins~\cite{Lieb61, Barouch1971, McCoy68}. 
On the contrary, a new type of order, which is due to geometrical frustration, is found in the region $\phi\in(-\pi/4,\pi/4)$, where the antiferromagnetic interactions dominate. 
Without frustration, this region would be simply a $x$-AFM phase characterized by a staggered magnetization. 
Instead, assuming FBC, it is separated into two gapless regions (the energy gap closing as $1/N^2$), $\phi\in(-\pi/4,0)$ and $\phi\in(0,\pi/4)$, characterized by different ground state degeneracies and different magnetization patterns. 
Moreover, the transition is accompanied by a finite discontinuity in the first derivative of the ground state energy at $\phi=0$.

For $\phi\in(-\pi/4,0)$, where the dominant antiferromagnetic interaction in the $x$ direction competes with the ferromagnetic one in the $y$ direction, the ground state manifold is two-fold degenerate. 
Although the dominant interaction along $x$ is antiferromagnetic, the magnetization $m_x(j)=\braket{\sigma_j^x}$ (as well as $m_y(j),m_z(j)$) is found to be uniform, ferromagnetic, and decays algebraically with the system size to zero, as $1/N$, resulting in the zero value of the magnetization in the thermodynamic limit. Qualitatively, this behavior stems from the fact that with an odd number of sites with periodic boundary conditions, a staggered order cannot be sustained, and thus the delocalized kink contribution eventually destroys the AFM order. Because of these properties, this order is termed \textit{Mesoscopic Ferromagnetic Order}~\cite{Maric20}.

For $\phi\in(0,\pi/4)$, where both interactions are antiferromagnetic, a more rich behavior is found. The ground state manifold is four-fold degenerate and it is possible to select ground states with different properties. While there are states that also exhibit mesoscopic ferromagnetic order, it is also possible to select states with a magnetization profile that varies in space with an incommensurate pattern and survives in the thermodynamic limit. Qualitatively, in this case, the system accommodates the frustration with a small shift in the staggered order, so that the magnetization varies as $\sin \left[ \pi \left(1 - \frac{1}{N} \right) j+ \alpha \right]$: neighboring sites are almost perfectly staggered, but along the chain, the amplitude varies and at its minimum one finds a ferromagnetic bond. This new type of order has been termed \textit{Incommensurate Antiferromagnetic Order}~\cite{Maric20_2}.

\section{Method of analysis}\label{Methods}

The model in eq.~\eqref{HamDef} can be analyzed by mapping spins into spinless non-interacting fermions through the Jordan-Wigner transformations~\cite{Lieb61, Jordan28}. 
Usually, in systems that can be solved exploiting the Jordan-Wigner transformation, followed by a Bogoliubov rotation in Fourier space, all the physical quantities can be obtained in terms of two-body correlation functions of Majorana operators that are determined analytically~\cite{Lieb61, Barouch1971, Smacchia2011, Giampaolo2015, Zonzo2018, Maric20_3}.
However, in the present case, the local perturbation explicitly breaks the invariance under spatial translation and, therefore, prevents the possibility to obtain the analytical expressions of the Majorana correlation functions.
Nevertheless, since the Hamiltonian in eq.~\eqref{HamDef}, is quadratic in the spinless fermion operators, we construct an efficient algorithm, based on the work of Lieb \textit{et al.}~\cite{ Lieb61} to obtain a numerical evaluation of the whole set of Majorana correlation functions that allows to obtain all the analyzed quantities following the standard approach  (see Appendix~\ref{NumAlg} for details). 

Usually, a finite longitudinal field is required to have a finite magnetization in the $x$-direction. 
The persistence of a finite value even after the removal of the field, after taking the thermodynamic limit, is the signature of a spontaneous symmetry breaking.
However, in our case, we are working at zero fields to have an exact degeneracy between states with different parities, so that the system can exhibit a finite magnetization, even at a finite size, without the need to apply a symmetry-breaking field.
Since the different parity operators do not commute with each other, any ground state vector necessarily breaks at least one of those symmetries.  
Once magnetizations are obtained for a chosen $N$, we follow this value toward the thermodynamic limit to determine which order survives for large systems.
Taking inspiration from the result obtained in the absence of defect~\cite{Maric20, Maric20_2}, we focus mainly on the study of the pattern of magnetization in the $x$ direction $m_x(j)=\braket{\sigma_j^x}$ which is maximized by taking into account one of the states with definite $\Pi^x$ parity that reads
\begin{equation}
 \ket{g}=\frac{1}{\sqrt{2}}(\mathbb{1}+\Pi^x)\ket{g^-} \;.
\end{equation}
where $\ket{g^-}$ is the ground state of the Hamiltonian in eq.~\eqref{HamDef} that falls in the odd sector of $\Pi^z$. 
Exploiting the trick introduced in~\cite{Maric20}, we can express the expectation value of $\sigma_i^x$ on $\ket{g}$ in terms of expectation value of the operator $\sigma_i^x \Pi^x$ on $\ket{g^-}$, i.e.
\begin{equation}
m_x(j)\equiv\bra{g}\sigma_j^x\ket{g}=\bra{g^-}\sigma_j^x  \Pi^x\ket{g^-}
\end{equation}
which can be computed using the fermionic representation of the model, as discussed in Appendix~\ref{NumAlg}.

To further corroborate these results, we also employed an analytical perturbation theory, in two different ways.
Treating either $\phi$ or $\delta_x$ in eq.~\eqref{HamDef} as a small parameter, we expanded either in the kink state basis or just in the four-dimensional ground state manifold on the unperturbed model.
Details are given in Appendix~\ref{appendix perturbation theory}. The first approach is more insightful and successful in describing the numerical results, while the second gives a more quantitative agreement for the incommensurate AFM order, although the truncation of the basis to just four states is only empirically justified.

With a finite defect, the thermodynamic limit presents an additional challenge, since the ground state manifold sits at the bottom of a band of $2N$ states whose density increases with the system size. The analysis in \cite{Campostrini2015} indicate that it is possible to scale the defect strengths $\delta_{x,y}$ with the system size to preserve the orders found with FBC (namely, in a way that $\delta_{x,y} \to 0$ as $N \to \infty$). However, we are interested in determining the resilience of the orders found in~\cite{Maric20, Maric20_2} to the presence of a finite defect in the thermodynamic limit. To do so, we fix the strength (in absolute value) of the defect to be comparable with the energy width of the (unperturbed) lowest energy band, namely $\vert\delta_{x,y}\vert = \vert\phi\vert/10$. In this way, the defect always hybridizes several states of the band proportional to $N$ such that the finite-size effects are under better control.

In our analysis, we will have to face several different magnetization patterns and, hence, we have to find a way to discriminate among them. 
Even if, sometimes, it would be enough to look to a direct plot of the magnetizations to guess what kind of pattern is realized in the system, it would be better to have a more quantitative way to discriminate between them. 
For this reason we decide to focus on the analysis of its Discrete Fourier Transform (DFT):
\begin{equation}\label{DFTDef}
\!\!\tilde{m}_x(k) \!\equiv\!
\dfrac{1}{N}\sum_{j=1}^N m_x(j)  e^{\frac{2\pi \imath}{N} k j},\; \; \; \; k\!=\!1,\ldots,N.
\end{equation}
Hence to determine the asymptotic behavior of the magnetization pattern in the thermodynamic limit, we will perform a finite size scaling analysis of the DFT, and we will compare the result so obtained with some reference patterns. 

For instance, the incommensurate AFM order has $\tilde{m}_x(k) \propto \delta_{k, \frac{N\pm 1}{2}}$, while the mesoscopic ferromagnetic order would have 
{$\tilde{m}_x(k) \propto \delta_{k,N}$}, but with an amplitude decaying algebraically to zero as $N \to \infty$. Finally, a perfectly staggered order would have $\pi$-momentum, which is, however, not allowed by the quantization rules with FBC. Thus, such order would be resolved over the allowed momenta as
\begin{equation}
\label{AFMDFT}
\tilde{m}_x(k) \propto \frac{1}{1 + e^{-\frac{2\pi \imath}{N}  k }} \; . 
\end{equation}

\section{Results}\label{Results}

\begin{figure}[t]
\includegraphics[width=1\linewidth]{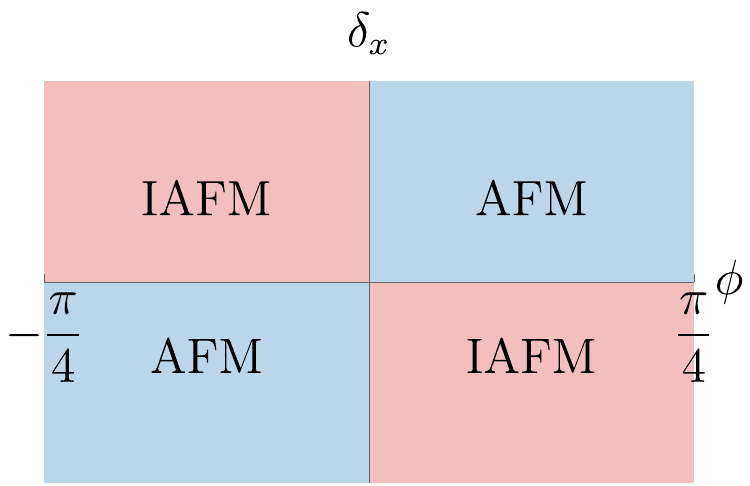}
\caption{{Schematic representation of the different phases as function of the system parameters.
The interval $(-\pi/4, \pi/4)$ corresponds to a dominating antiferromagnetic interaction along the $x$ axis
When a defect of the form $(\delta_x \neq 0, \delta_y = 0)$
favours a ferromagnetic alignment, namely when a FTD defect is considered, the effect of the defect is localized in a small region and outside of them the standard AFM order is restored.
On the other hand when it tends to strengthen the antiferromagnetic interaction (AFTD) the incommensurate AFM order (IAFM) is realized. }}\label{fig:phasesrepr}
\end{figure}

\begin{figure*}[t]
	\begin{subfigure}{0.49\textwidth}
		\includegraphics[width=0.95\linewidth]{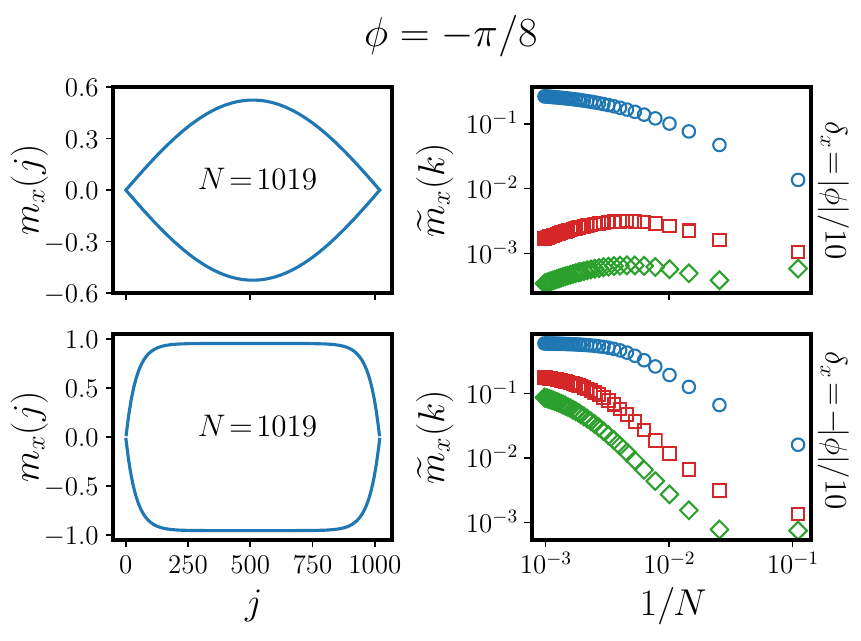}
		\caption{}
	\end{subfigure}
	\begin{subfigure}{0.49\textwidth}
		\includegraphics[width=0.95\linewidth]{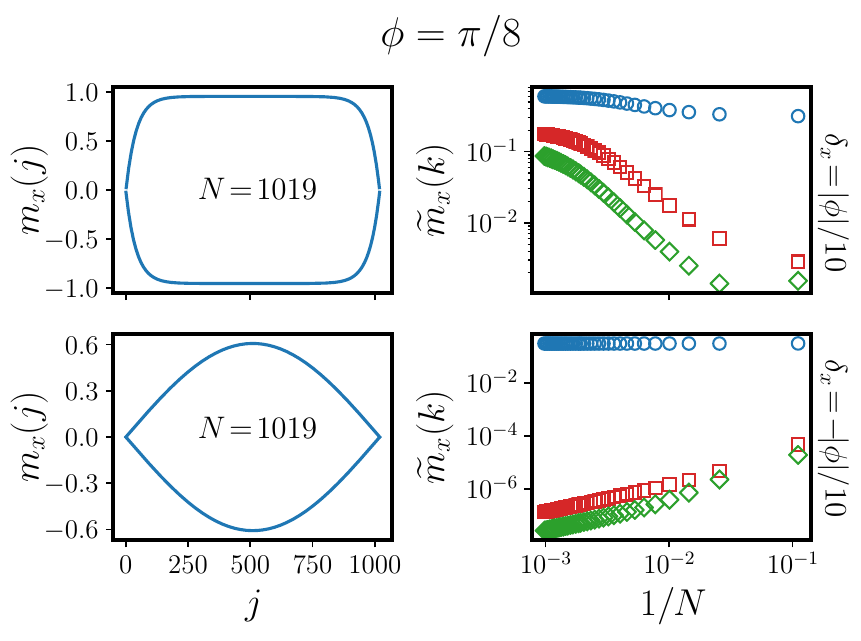}
		\caption{}
	\end{subfigure}
	\caption{Magnetization $m_x (j)=\braket{\sigma_j^x}$ as a function of the site $j$ for a chain made of $N=1019$ spins (left), and the absolute value of its Discrete Fourier transform (DFT) eq.~\eqref{DFTDef} as a function of the inverse chain length, for chain lengths up to $N=2001$, (right) for different momenta: green diamonds $\vert\widetilde{m}_x(\frac{N\pm5}{2})\vert$, red squares $\vert\widetilde{m}_x(\frac{N\pm3}{2})\vert$, and blue circles $\vert\widetilde{m}_x(\frac{N\pm1}{2})\vert$. The results are obtained considering the defect only along the $x$ direction ($\delta_y=0$). An antiferromagnetic defect yields an incommensurate AFM order, while a ferromagnetic one gives standard AFM order in the bulk (see text for discussion).}
	\label{fig:1}
\end{figure*}

We can now start to illustrate the results of our analysis of the behavior of the magnetization $m_x(j)$ under the presence of a defect for finite chains and then extrapolate its behavior in the thermodynamic limit in the two frustrated regions studied in~\cite{Maric20, Maric20_2}.
At first, we focus on the case where the defect affects only one spin direction, and then we switch to the case where the defect acts on both. 

{Before starting a detailed discussion, in Fig.~\ref{fig:phasesrepr} we depict a schematic phase diagram of the system in the presence of a defect with $(\delta_x\neq0,\delta_y=0)$.
In it we can see that the system can show two clearly different behavior.
As we can see from Fig.~\ref{fig:1},} when the defect tends to strengthen the AFM interactions, i.e. when an AFTD defect is considered, the max of the DFT, that is obtained for $k=\frac{N\pm1}{2}$ goes to a non-zero value as the system size diverges while $\tilde m_x(k)$ vanishes for all other momenta. 
This picture is coherent with an incommensurate AFM order in which the site-dependent magnetization is proportional to $\sin \left[ \pi(1\!-\!\frac{1}{N}) j \right]$ as can also be seen from the plots of the envelopes obtained for $N=1019$ spins.
It is worth noting that the incommensurate AFM order is found both for $\phi\in(0,\pi/4)$ and for $\phi\in(-\pi/4,0)$, although in the latter region, without perturbation, a mesoscopic ferromagnetic order was present. Thus an AFTD stabilizes the incommensurate AFM order, regardless of the order that characterizes the unperturbed underlying model.

A peculiar feature needs commenting: Although one could naively expect that a stronger AFM bond would concentrate around the defect the largest magnetization amplitude, this is not the case and one observes the magnetization minimum at the defect for both signs of the latter. We do not have a satisfactory qualitative explanation for this behavior, although the perturbative calculations below provide some technical justifications. It seems that the system prefers to have the most constant magnetization profile far away from the defect, so that at the center of the chain the order is hardly distinguishable from the unfrustrated one. Although the reaction to FBC is to excite a single quasi-particle over the vacuum, we cannot characterize the observed position of the magnetization minimum as anything else but a many-body effect.

\begin{figure}[t]
     \includegraphics[width=1\linewidth]{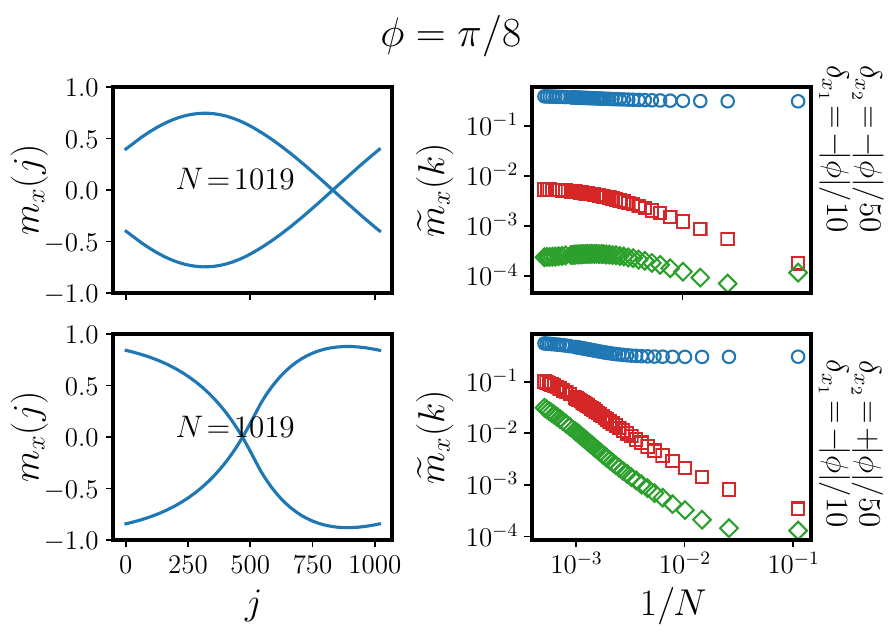}
	\caption{Magnetization with two bond defects, one between the first and the last spins of the chain $(\delta_{x_1})$ and the second between the $\frac{N-1}{2}$ and the $\frac{N+1}{2}$ spins  $(\delta_{x_2})$. On the left, the magnetization $m_x (j)=\braket{\sigma_j^x}$ as a function of the site $j$ for $N=1019$ and on the right the absolute value of its Discrete Fourier transform (DFT) in eq.~\eqref{DFTDef} as a function of the inverse chain length, up to $N=2001$, for different momenta: green diamonds $\vert\widetilde{m}_x(\frac{N\pm5}{2})\vert$, red squares $\vert\widetilde{m}_x(\frac{N\pm3}{2})\vert$, and blue circles $\vert\widetilde{m}_x(\frac{N\pm1}{2})\vert$. The results are obtained considering the defects only along the $x$ direction $(\delta_{y_1}= \delta_{y_2}=0)$. While in the case of both defect being AFTD (upper row) the DFT clearly signals the rising of a macroscopic incommensurate staggerization in which the magnetization on the $j$-th spin is proportional to $\sin \left[ \pi(1\!-\!\frac{1}{N}) j \right]$, when the smaller defect is FTD the system sizes considered are not sufficient to clearly characterize the emerging order. It seems clear that at least one Fourier component remains finite as $N \to \infty$, indicating that the order survives in the thermodynamic limit, but the determination of the faith of the other components requires larger $N$'s.} 
	\label{fig:2def}
\end{figure}

As we mentioned above, a single bond defect breaks all the mirror symmetries of the chain, except the one crossing the site $\frac{N+1}{2}$. Accordingly, the magnetization pattern with an AFTD satisfies this mirror symmetry and one can wonder how much this fact constraints its regular structure. Hence, a question that arises naturally is if the incommensurate AFM pattern survives even when a second localized defect is added to the Hamiltonian in eq.~\eqref{HamDef} to also breaks the remaining mirror symmetry. Thus, we introduce a smaller bond defect between the $\frac{N-1}{2}$-th and the $\frac{N+1}{2}$-th spins and present our results for this case Fig.~\ref{fig:2def}.
Due to the second defect, the convergence of the DFT is quite slow and chains longer than $N=2001$ would be required to clearly reach the asymptotic behavior. 
Nonetheless, the max of the DFT, obtained for $k=\frac{N\pm1}{2}$ indicates that the incommensurate staggerization always survives in the thermodynamic limit. When both defects are AFTD, it seems that once more an incommensurate AFM order is established, proving that its existence is not dependent on the presence of a mirror symmetry which needs to be respected. With the weaker defect being FTD, it is not clear what will happen in the thermodynamic limit, but the pattern is not qualitatively too different from before. All these results indicate that, even under realistic experimental situations (i.e. without perfect translational invariance), the incommensurate AFM can exist and be observed. As a side note, while the defects are placed on the same bonds in both cases, with two AFTDs, the minimum of the magnetization is located at an intermediate point between them, while the FTD fixes it very close to itself, but not on it, probably because of a finite correlation length.

\begin{figure}[t]
     \includegraphics[width=1\linewidth]{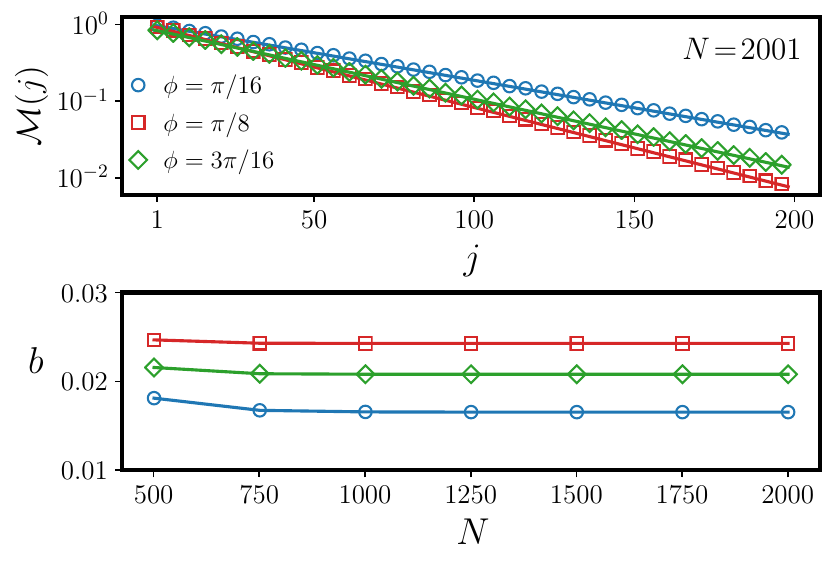}
	\caption{Upper panel. Behavior of $\mathcal{M}(j)$ as function of $j$ for a system made of 2001 spins for different value of $\phi$.  Bottom panel. Size dependent numerical extimation of the exponential ratio $b$ for different value of $\phi$. In both panel the blue circle stands for results obtained settings $\phi=\frac{\pi}{16}$, the green diamond for $\phi=\frac{\pi}{8}$ and the red square for $\phi=\frac{3 \pi}{16}$.}
	\label{fig:new}
\end{figure}

Turning back to the case of a single defect, the picture changes abruptly when the defect turns to be an FTD, i.e. when it starts to suppress the dominant antiferromagnetic interaction on one bond.
Not only the maximum but all the $\widetilde m_x(k) $ go towards finite values, satisfying precise ratio rules, such as $\frac{\widetilde m_x(\frac{N\pm3}{2})}{\widetilde m_x(\frac{N\pm1}{2}
)}=\frac{1}{3}$,$\frac{\widetilde m_x(\frac{N\pm5}{2})}{\widetilde m_x(\frac{N\pm1}{2})}=\frac{1}{5}$ etc.
This behavior is compatible with a perfectly staggered AFM order in the bulk, with deviations localized around the defect. As mentioned above, such bulk behavior would be characterized by a sharp peak at $\pi$, corresponding to a wavenumber $N/2$, which, being $N$ odd, is not allowed. The aforementioned ratios can be easily obtained by taking the $N \to \infty$ limit of eq.~\eqref{AFMDFT} and reflect the expansion of a perfect AFM order over the available wavenumbers, which are symmetrically distributed around $N/2$.
As we can see also from the envelopes, the region affected by the presence of the defect is small because its effect decays exponentially.
This fact can be better appreciated by looking at Fig.~\ref{fig:new} where we have depicted the behavior of the function
\begin{equation}
 \mathcal{M}(j)=\max_l|m_x(l)|-|m_x(j)|,
\end{equation}
where $\max_j|m_x(j)|$ represents the value of the magnetizations in the bulk and the analysis of $\mathcal{M}(j)$ helps to understand the dimension outside which the effect of the defect is suppressed. 
As we can see from the figure, the effect decays exponentially $\mathcal{M}(j,\phi)\propto e^{-bj}$ at an exponential ratio $b$ that, for large $N$ becomes size independent. 
Hence in the thermodynamic limit, the magnetic pattern is similar to the one of a kink state, where the latter is localized at a distance $b$ around the defect, and away from it tends to the standard AFM staggerization.

However, Fig.~\ref{fig:1} shows also another important result. 
By keeping the $\delta_x$ fixed and changing $\phi$ from $\frac{\pi}{8}$ to $-\frac{\pi}{8}$ or by fixing the value of $\phi\in(-\frac{\pi}{4},\frac{\pi}{4})$ and changing the defect from AFM to ferromagnetic (or vice-versa),
we have an abrupt change of the magnetization pattern. At one side we have a standard AFM order with a localized defect and on the other an incommensurate staggerization.
The existence of two different orders is compatible with the results by  Campostrini \textit{et al.}~\cite{Campostrini2015}, although they did not consider the behavior of the local order: Changing the defect from AFM to ferro (and viceversa) in a chain with FBC indeed drives the system across a QPT. However, our case is more rich than in~\cite{Campostrini2015}, since in our model we also cross a QPT by varying the bulk interaction parameter $\phi$, with and without the defect.
In any case, the most important point is that these quantum phase transitions
cannot exist without FBC, as we have verified by cutting the lattice far away from the position of the defect (between the spin at $\frac{N-1}{2}$ and the one at $\frac{N+1}{2}$): in this case, the incommensurate order is not supported and in the whole region the system realizes a ground state with a standard AFM staggerization plus a localized defect.
Hence, once more we see that FBC provide a challenge against the standard tenants of Landau's Theory.

\begin{figure}[b]
	\includegraphics[width=1\linewidth]{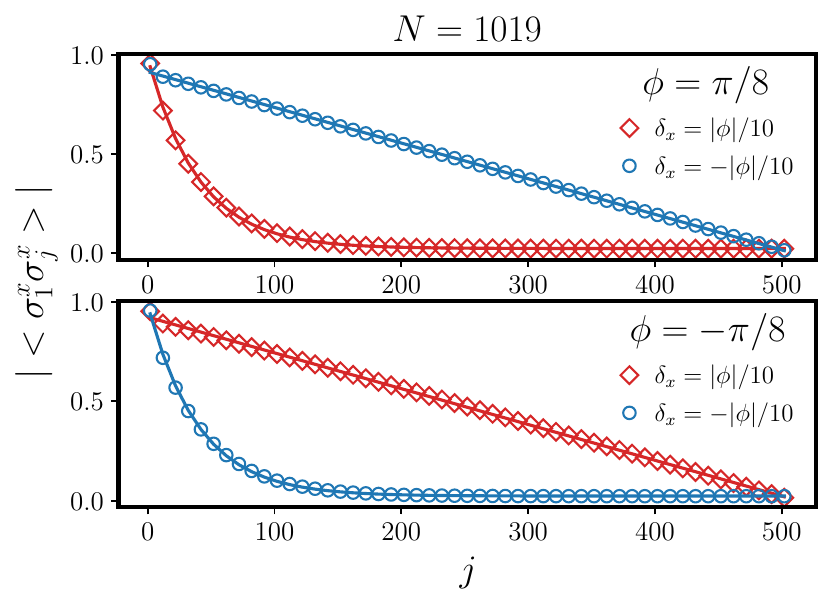}
	\caption{{Spin correlation $|<\sigma^x_1\sigma^x_j>|$ as a function of the site $j$, for a chain made of $N=1019$ spins. The defect is along the $x$ direction. The exponential decay found for a FTD defect underlines the standard AFM order. On the other hand the linear decay for the AFTD case is a typical signature of the frustrated nature of the incommensurate AFM phase.}}
	\label{fig:corrFun}
\end{figure}

{We finally note how the different natures of the two magnetic orders are also highlighted by the analysis of the spatial dependence of the spin correlation functions. 
Indeed, in Fig.~\ref{fig:corrFun}, we analyze the spin-spin correlation function along the $x$ spin direction $\braket{\sigma^x_1\sigma^x_j}$ as a function of $j$. 
The localization of the effect of the defect that characterize the FTD is visible in the exponential decay $|\braket{\sigma^x_1\sigma^x_j}| = a+b e^{-c j}$.
On the other hand the behaviour $|\braket{\sigma^x_1\sigma^x_j}| = a- b(j-1)$, that is a typical signature of the frustrated phase, is observed for the AFTD.}

We can get more insight and reach the same conclusions, about the different magnetic order, through a perturbative analysis. We have done two different perturbation theories, which are presented in Appendix~\ref{appendix perturbation theory}. In both approaches, we are going to ignore every state separated by a finite energy gap from the ground states. However, the ground states are part of a band of $2N$ states for which, in the thermodynamic limit, the gap vanishes, complicating any perturbative calculation. Thus, in our first approach, we worked around the point $\phi=0$, and in this way, we provide a good picture explaining our numerical results. 
At $\phi=0$ the ground state manifold is $2N$-fold degenerate, spanned by the ``kink'' states which have a ferromagnetic bond $\sigma_j^x= \sigma_{j+1}^x=\pm 1$ and AFM bonds between all other sites, for $j=1,2,\ldots N$. Adding the small interaction in the $y$ direction, proportional to $\sum_j\sigma_j^y\sigma_{j+1}^y$, the kink states split in energy. By developing a method introduced in \cite{Campostrini2015_2}, we are then able to diagonalize the band of kink states under this term and do not need to deal with the complications emerging from a perturbative series with closing energy gaps. In the case of FTD, the ground states are, to the lowest order in the perturbation theory, simply the kink states with the ferromagnetic bond between the first and last site ($\sigma_1^x=\sigma_N^x=\pm 1$), and the other states are separated by a finite energy gap determined by $\delta_x$. In the case of AFTD, the ground states are superpositions of kink states that have  $\sigma_1^x=-\sigma_N^x$ and they belong to a band of states, in which the energy gap between the states closes as $1/N^2$, as in frustrated models without the defect \cite{Maric20, Maric20_2, Giampaolo2019, Dong2016}. Both the two cases are characterized by a two-fold degenerate ground-state manifold, as expected.

Having the ground states, to the lowest order in perturbation theory, the magnetization can be computed. In the case of an FTD, we find that for both signs of $\phi$ the magnetization is given by
\begin{equation}
m_x(j)=(-1)^j,
\end{equation}
which represents a standard AFM staggerization, apart from the ferromagnetic bond placed where the defect is. The numerical results of Fig.~\ref{fig:1} show indeed standard staggered magnetization far from the defect, but zero value where the defect is placed. Thus the perturbation theory explains well the bulk behavior of the system, far from the defect. Close to $\phi=0$ the correlation length is small and the kinks are extremely well localized, while the numerics refer to a choice of a finite correlation length that provides a length scale over which the presence of the defect is felt before the bulk order ensue (see Fig.~\ref{fig:new}). 

In the case of an AFTD, the perturbation theory predicts, for both signs of $\phi$, the magnetization
\begin{equation}\label{incommensurate perturbation finite size}
  m_x(j)=\frac{(-1)^j\sin\big[\frac{\pi}{N}\big(j-\frac{1}{2}\big)\big]}{N\sin\big(\frac{\pi}{2N}\big)}+\frac{1}{N},
\end{equation}
which for large $N$ is well approximated by
\begin{equation}\label{magnetization result tl phi zero}
  m_x(j)=(-1)^j\frac{2}{\pi}\sin\Big(\frac{\pi}{N}j\Big).
\end{equation}
This order is the incommensurate AFM order found in \cite{Maric20_2}, with a locally staggered magnetization, but modulated in magnitude over the length of the chain. Against naive expectations, but in agreement with the numerics, the modulation is such to have an exactly vanishing magnetization at the sites connected by the defect, even when the latter would lower the energy of a strong AFM order. This perturbative calculation validates well our numerical results of Fig.~\ref{fig:1}, both close and far from the defect.

\begin{figure}[t]
     \includegraphics[width=1\linewidth]{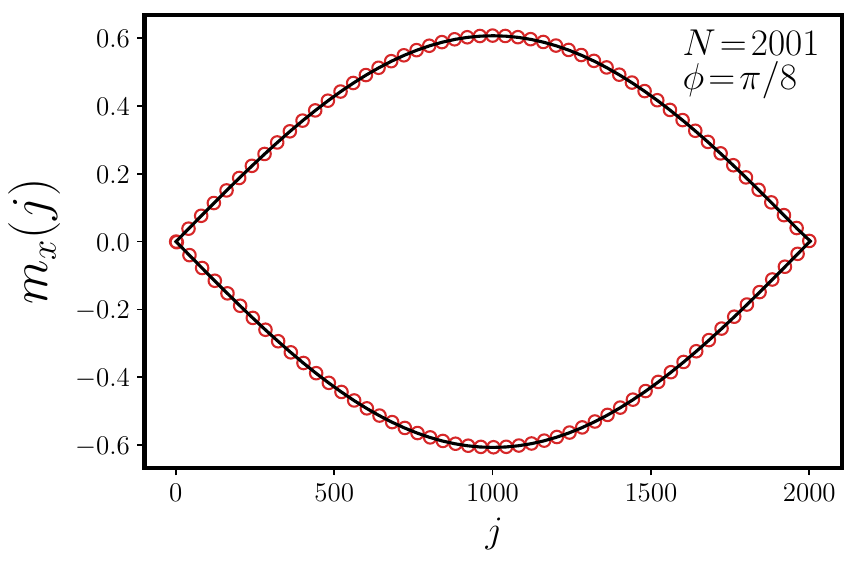}
	\caption{Comparison with the expression of the magnetization in eq.~\eqref{strange_mag} obtained with the perturbative approach (solid black line) and the data obtained with the exact calculation (red dots) for a system made by $N=2001$ spins and $\phi=\frac{\pi}{8}, \delta_x = \frac{\pi}{80}$.}
	\label{fig:comparison}
\end{figure}

\begin{figure*}[t]
   	\begin{subfigure}{0.49\textwidth}
		\includegraphics[width=1.\linewidth]{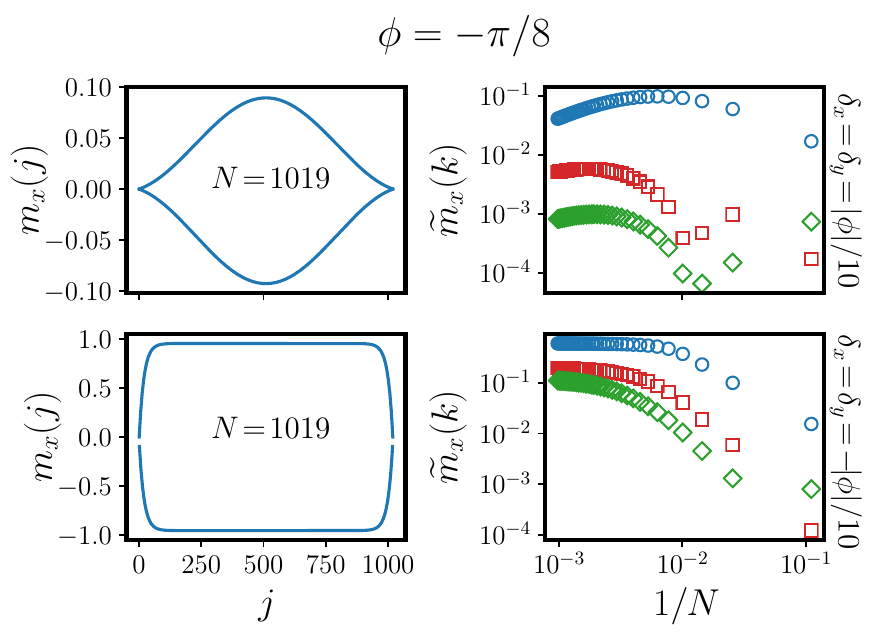}
		\caption{}\label{fig new order a}
	\end{subfigure}
	\begin{subfigure}{0.49\textwidth}
		\includegraphics[width=1.\linewidth]{both_pi_8}
		\caption{}\label{fig new order b}
	\end{subfigure}
	\caption{Magnetization $m_x (j)=\braket{\sigma_j^x}$ as a function of the site $j$ for a chain made of $N=1019$ spins (left), and the absolute value of its Discrete Fourier transform (DFT) in eq.~\eqref{DFTDef} as a function of the inverse chain length (right), while keeping fixed the ``momentum'': green diamonds $\vert\widetilde{m}_x(\frac{N-3}{2})\vert$, red squares $\vert\widetilde{m}_x(\frac{N-1}{2})\vert$, and blue circles $\vert\widetilde{m}_x(\frac{N+1}{2})\vert$. The defect is along both the $x$ and $y$ directions. An antiferromagnetic one yields mesoscopic magnetization that varies in space with an incommensurate pattern, while a ferromagnetic defect gives standard AFM order in the bulk (see text for discussion).}
	\label{fig:3}
\end{figure*}

The incommensurate AFM order is found whenever the defect is AFM, even though, without the defect, it is present only for $\phi\in(0,\pi/4)$ \cite{Maric20_2}.
Without the defect, this region possesses a four-fold degenerate ground-state manifold out of which it is possible to select the ground states exhibiting incommensurate AFM order. It is thus of interest to find which of these ground states are selected through a small antiferromagnetic defect. We answer this question in Appendix~\ref{sec perturbation 2}, by doing a (degenerate) perturbation theory for small $\delta_x$. In this case, we ignore the band above the ground states and considered the effect of the defect only on the GS manifold. Of course, the resulting lowest energy state is odd under the mirror symmetry passing across the site $\frac{N+1}{2}$. Combining the obtained ground state with the techniques developed in~\cite{Maric20_2}, for the magnetization in the thermodynamic limit we find,
\begin{equation}
\label{strange_mag}
m_x(j)=(-1)^j\frac{2}{\pi}(1-\tan^2\phi)^{1/4}\sin\Big(\frac{\pi}{N}j\Big),
\end{equation}
which generalizes eq.~\eqref{magnetization result tl phi zero} to the whole region $\phi\in(0,\pi/4)$ and is in good agreement with our numerical results, see Fig.~\ref{fig:comparison}. Note that in the perturbation theory in $\delta_x$ we have neglected all excited states of the unperturbed model, including those belonging to the lowest energy band. While in the case of interest the procedure yields a result in agreement with numerics, this approach is not justified in general, because of the gapless nature of the unperturbed system. While the quantitative agreement of this approach with the numerics is quite surprising, it provides a geometrical explanation of the observed qualitative behavior. In fact, since the defect preserves the mirror symmetry across the site $\frac{N+1}{2}$, states (which always come in degenerate duplet or quadruplets) are hybridized to select the combination in each multiplet with definite mirror symmetry (with the odd one having lower energy): By explicit construction we observe that both states even or odd under mirror symmetry have vanishing magnetization at the defect, in the thermodynamic limit. At finite size, the former have a small ferromagnetic bond at the defect, while the latter exhibit exactly zero value of the magnetization at the defect.

While we did not find qualitative differences between turning on a defect either in the $x$ or in the $y$ direction, the same cannot be said when both are finite.
We are now in presence of two defects which can agree or disagree in favoring a ferromagnetic or an anti-ferromagnetic alignment along with that bond in either direction. 
A typical example of our numerical results for these cases is given in Fig.~\ref{fig:3}. 
In the case when both defects suppress the dominant antiferromagnetic interaction, we can see a picture completely analogous to the case with a defect equal zero and the second acting as an FTD.
The behavior of the DFT is compatible with the Fourier Transform of a single kink state and the real space magnetization envelope shows the effect of the defect of decay exponentially fast moving away from it to reach a regular AFM pattern in the bulk.
Hence in the thermodynamic limit, the magnetic pattern is completely equivalent to the one of a single kink state that is, except a few sites around the main defect, the same of the standard AFM staggerization.

On the contrary, when both the defect are AFTD, all the elements of the  
DFT go very slowly to zero in the thermodynamic limit, thus signaling that the magnetization pattern is mesoscopic, i.e. that all magnetizations vanish in the thermodynamic limit. 
As in the case in which $\delta_y$ was set to zero, also in this case we can see that either by keeping fixed the defect and changing $\phi$ from $\frac{\pi}{8}$ to $-\frac{\pi}{8}$, or by fixing $\phi$ and turning the defect from AFM to ferro (or viceversa), we have that the magnetization pattern changes abruptly from the standard AFM one with a localized defect to one with an incommensurate mesoscopic staggerization. Hence, also in this case, the presence of these two different magnetization patterns signals the presence of a QPT, induced by the FBC.
The fact that this transition is no more present if we open the chain, by cutting the interaction between two neighboring spins, is further proof of the resilience of frustration effects to the presence of defects.

\section{Conclusions and Outlook}\label{Conclusions}

We have analyzed a generalization of the model previously studied in~\cite{Maric20, Maric20_2}, i.e. the XY chain at zero fields with FBC to which a localized defect in the Hamiltonian has been added. 
The resulting system has been characterized through the behavior, both at finite-sizes and in the thermodynamic limit, of the magnetization $m_x (j)=\braket{\sigma_j^x}$. Our motivation has been to challenge the na\"ive expectation that, being the effect of FBC to ignite a single excitation into the system, a defect would immediately spoil the features discovered in~\cite{Maric20, Maric20_2}.

We have found that depending on the kind of defects we add, the system responds by selecting different types of orders. 
According to expectations, a defect in the Hamiltonian that reduces the relative weight of the dominant antiferromagnetic interaction forces a ferromagnetic alignment along with the bond with the defect, resolving the frustration and restoring the standard AFM order except for a small region around the defect. 
On the contrary, if the bond defect favors an AFM alignment, we find an incommensurate AFM order: Locally, two neighboring spins are anti-aligned but, along the chain, the amplitude of the magnetization varies in space and vanishes at the defect. For a defect aligned along either the $x$ or the $y$ axis, this order is the same originally discovered in~\cite{Maric20_2} and survives the thermodynamic limit, while, when a defect in both directions is considered, the envelope of the magnetization changes from a sine to a sine-squared and its amplitude vanishes as $N \to \infty$.

While the resurgence of the traditional AFM order is in line with the traditional expectation that FBC can be accounted for by single particle physics, the other orders challenge this point of view and promote a more many-body interpretation, as signaled by the fact that the largest amplitude of the magnetization is not placed at the AFM defect.

All this outcome strengthens the idea in~\cite{Maric20, Maric20_2}, that FBC can induce a QPT which is absent if other boundary conditions are considered and that across this QPT a change in the local order can be detected.
Furthermore, these results corroborate the analysis in~\cite{Campostrini2015}, indicating that translational invariant system with FBC lie at the transition between a magnetic phase for a ferromagnetic defect that restores the un-frustrated order, and a kink phase when the defect is AFM and the local order remains sensitive to subdominant contributions.
Since FBC are the threshold for a QPT, we should not be surprised by the sudden change in the local order driven by a defect: Indeed, at finite sizes, one could scale the defect strength in various ways with the system size to follow the emergence of new orders, compared to the pure FBC case, but our emphasis is on what happens in the thermodynamic limit. In this way, we have shown that translational invariance is not a necessary condition for the appearance of frustrated phases, paving the way to its experimental observability and demonstrating that, close to FBC, the standard AFM order is not generically stable.

The renewed interested in the study of the FBC is yet at the beginning and can be expanded in various directions. An undoubtedly interesting research area is the study of the influence of the FBC on various types of orders such as the topological and nematic ones. A first step was made in~\cite{Maric20_3}, where some evidence has been collected suggesting that topological orders are resilient to topological frustration. Regarding the presence of defects in the chain, more complex situations can be considered by varying both the type of interaction and the number of defects. These analyses will be the topic of future research.

\section*{Acknowledgments}
We acknowledge support from the European Regional Development Fund -- the Competitiveness and Cohesion Operational Programme (KK.01.1.1.06 -- RBI TWIN SIN) and from the Croatian Science Foundation (HrZZ) Projects No. IP--2016--6--3347 and IP--2019--4--3321.
SMG, FF and {GT} also acknowledge support from the QuantiXLie Center of Excellence, a project co--financed by the Croatian Government and European Union through the European Regional Development Fund -- the Competitiveness and Cohesion (Grant KK.01.1.1.01.0004).

\appendix

\section{Numerical procedure}\label{NumAlg}

To diagonalize the Hamiltonian in eq.~\eqref{HamDef} we resort to the Jordan-Wigner transformation~\cite{Jordan28,FranchiniBook}, that maps spin operators into fermionic ones:
\begin{equation}\label{JWTransf}
	c_j=\left(\bigotimes_{k=1}^{j-1}\sigma_k^z\right)\otimes\sigma_j^+,\qquad c_j^\dagger=\left(\bigotimes_{k=1}^{j-1}\sigma_k^z\right)\otimes\sigma_j^-,
\end{equation}
where $\sigma_j^\pm=(\sigma_j^x\pm\imath\sigma_j^y)/2$ are the Pauli ladder operators. Through eq.~\eqref{JWTransf} the Hamiltonian in eq.~\eqref{HamDef} can be recast into the form:
\begin{eqnarray}\label{HamFerm}
H\!&=&\!\sum_{j=1}^{N-1}\! \left[\left(\cos\phi\!-\!\sin\phi\right)c_j c_{j+1}\!-\!\left(\cos\phi\!+\!\sin\phi\right)c_j c_{j+1}^\dagger\right]\!+\! \nonumber \\
& & -\Pi^z\Big[\left(\cos(\phi+\delta_x)\!-\!\sin(\phi+\delta_y)\right)c_Nc_{1}\!+\Big. \nonumber \\
& & \Big.-\!\left(\cos(\phi+\delta_x)\!+\!\sin(\phi+\delta_y)\right)c_Nc_{1}^\dagger\Big]\!\!+\textrm{h.c.}
\end{eqnarray}
Since $[H,\Pi^z]=0$ we can identify two different disconnected sectors corresponding respectively to the values $\Pi^z=\pm 1$. In the following we focus on the Hamiltonian of the odd sector, i.e. we fix $\Pi^z=-1$, since, once obtained the ground state $\ket{g^-}$ in this case, the other one with the same energy in the even sector is $\Pi^x\ket{g^-}$, up to a phase factor.

The Hamiltonian in eq.~\eqref{HamFerm} is quadratic in the fermionic operators, i.e. it can be rewritten as:
\begin{equation}\label{QuadForm}
	H\!=\!\sum_{j=1}^{N-1}\!\left[c_j^\dagger A_{j,j+1}c_{j+1}\!+\!\dfrac{1}{2}\left(c_j^\dagger B_{j,j+1}c_{j+1}^\dagger+\textrm{h.c.}\right)\right],
\end{equation}
where the matrices $A=A^\dagger$ and $B^\dagger =-B$ can be easily obtained by inspection from eq.~\eqref{HamFerm}. Following standard techniques~\cite{Lieb61} we introduce the linear transformation:
\begin{align}
&\eta_k=\sum_i\left[\dfrac{\Phi_{ki}+\Psi_{ki}}{2}c_i+\dfrac{\Phi_{ki}-\Psi_{ki}}{2}c_i^\dagger\right] \\
&\eta_k^\dagger=\sum_i\left[\dfrac{\Phi_{ki}+\Psi_{ki}}{2}c_i^\dagger+\dfrac{\Phi_{ki}-\Psi_{ki}}{2}c_i\right],
\end{align}
where the vectors $\Phi_k$ and $\Psi_k$ are given by the solution of the eigenvalue problems:
\begin{align}
	&\Phi_k(A-B)(A+B)=\Lambda_k^2\Phi_k \label{EigEq1} \\
	&\Psi_k(A+B)(A-B)=\Lambda_k^2\Psi_k. \label{EigEq2}
\end{align}
The Hamiltonian in eq.~\eqref{QuadForm} can then be reduced to the form:
\begin{equation}
	H=\sum_k\Lambda_k\eta_k^\dagger\eta_k+\dfrac{1}{2}\left[\Tr A-\sum_k \Lambda_k\right],
\end{equation}
where we define the energies $\Lambda_k$ to be all positive.

At variance with the unperturbed model, in which the GS degeneracy depends on the type of interaction, tuned by the $\phi$ parameter, the GS of the perturbed one is two-fold degenerate, due to the breaking of the translational invariance of the system. Furthermore, as discussed in the main text, since the Hamiltonian commutes with the $\Pi_z$ operator ($[H,\Pi_z]=0$), the  most general GS is of the form~\cite{Maric20,Maric20_2}:
\begin{equation}\label{gs}
\ket{g}=\left(\cos\theta+e^{i\psi}\sin\theta\;\Pi^x\right)\ket{g^-},
\end{equation}
where $\ket{g^-}$ is the (unique) GS of the system in the odd parity sector.

The magnetization for the ground state in eq.~\eqref{gs} is given by
\begin{equation}
\bra{g}\sigma_j^x\ket{g}=\cos\psi\sin(2\theta)\bra{g^-}\sigma_j^x\Pi^x\ket{g^-},
\end{equation}
since the matrix elements of $\sigma_j^x$ between different $\Pi^z$ sectors vanish. Of interest is the maximal magnetization that can be obtained on the ground state manifold. It is achieved in the states with definite $\Pi^x$ parity. Thus we have
\begin{equation}
m_x(j)=\expval{\sigma_j^x\Pi^x}{g^-} \label{magX2}
\end{equation}
for $\Pi^x=1$ (achieved by $\psi=0, \ \theta=\pi/4$) and
\begin{equation}
m_x(j)=-\expval{\sigma_j^x\Pi^x}{g^-}
\end{equation}
for $\Pi^x=- 1$ (achieved by $\psi=0, \ \theta=-\pi/4$). These are the magnetizations that we discuss in the main text.

Eq.~\eqref{magX2} can be evaluated expressing the operators on the r.h.s. in terms of Majorana fermions:
\begin{align}
&A_j=c_j^\dagger+c_j=\left(\bigotimes_{l=1}^{j-1}\sigma_l^z\right)\otimes\sigma_j^x, \\
&B_j=i(c_j^\dagger-c_j)=\left(\bigotimes_{l=1}^{j-1}\sigma_l^z\right)\otimes\sigma_j^y.
\end{align}
Furthermore we can resort to Wick theorem to express the expectation values in eq.~\eqref{magX2} in terms of the contractions $F(i,j)=-i\expval{A_iB_j}{g^-}$.

Let us denote the vacuum state for fermions $\eta_j$ by $\ket{0^-}$, i.e. we have $\eta_j\ket{0^-}=0$ for $j=1,2,\ldots N$. {We numerically verify, by direct computation,} that the parity of the state $\ket{0^-}$ is $\Pi^z=1.$ { On the other hand,} the Hamiltonian without defects is also written in terms of free fermions with positive energy \cite{Maric20_2} and the vacuum $\ket{0^-}$ has positive parity by construction there. 

Assuming the eigenvalue of the matrix appearing on the l.h.s. of eq.~\eqref{EigEq1} are labeled in descending order, the GS is then $\ket{g^-}=\eta_N^\dagger\ket{0^-}$. From this identification a straightforward calculation gives:
\begin{equation}
F(j,l)=-\imath \sum_{k=1}^{N-1}\Psi_{kj}\Phi_{kl}+\imath \: \alpha\Psi_{Nj}\Phi_{Nl},
\end{equation}
where $\alpha=\textrm{sgn}\left(\det A\right)$.

\section{Perturbation theory}\label{appendix perturbation theory}

In this section we study perturbatively the Hamiltonian in eq.~\eqref{HamDef} with $\delta_y=0$. Let us for the purpose of perturbation theory write the Hamiltonian as
\begin{equation}\label{Hamiltonian for perturbation theory}
H=\cos\phi\sum_{j=1}^{N}\sigma_j^x\sigma_{j+1}^x+\sin\phi\sum_{j=1}^{N}\sigma_j^y\sigma_{j+1}^y +\zeta \sigma_N^x\sigma_1^x,
\end{equation}
so that $\zeta>0$ corresponds to an antiferromagnetic defect, while $\zeta<0$ is a ferromagnetic defect. The case $\zeta=0$, of course, corresponds to FBC.

First, in section~\ref{sec perturbation 1} we are going to make the perturbation theory close to the classical point $\phi=0$, which explains well our numerical results. Then, in section~\ref{sec perturbation 2} using the perturbation theory around $\zeta=0$ we are going to find which of the four-fold degenerate ground states of the region $\phi\in(0,\pi/4)$, present without the defect, are selected by taking the limit of the small antiferromagnetic defect $\zeta\to 0^+$, which also explains well the order we have found numerically.

\subsection{Perturbation theory around $\phi=0$}\label{sec perturbation 1}

The perturbation theory around \mbox{$\phi=0$} without the defect (for $\zeta=0$) has already been done in \cite{Maric20,Maric20_2}. Without the defect, exactly at the classical point \mbox{$\phi=0$} the ground state manifold is $2N$-fold degenerate and consists of kink states \begin{equation}
\ket{j}=\ket{...,1,-1,1,1,-1,1,....} ,
\end{equation}
for $j=1,2\ldots N$, which have one ferromagnetic bond $\sigma_j^x=\sigma_{j+1}^x=1$ and antiferromagnetic bonds between other adjacent sites, and the kink states obtained from $\ket{j}$ by reversing all spins, which have $\sigma_j^x=\sigma_{j+1}^x=-1$ and all the other bonds antiferromagnetic. The latter can be written as $\Pi^z\ket{j}$. Note that the states $\ket{j}$ have the parity $\Pi^x=(-1)^{(N-1)/2}$, while $\Pi^z\ket{j}$ have, of course, the opposite parity. By turning on $\phi\neq 0$, the term proportional to $\sum_j\sigma_j^y\sigma_{j+1}^y$ kicks in and the $2N$-fold degenerate ground state manifold splits, resulting in the two-fold ground state degeneracy for $\phi<0$ and four fold for $\phi>0$.

The new ground states and the corresponding energies are found by diagonalizing the perturbation in the basis of the kink states, while the other states are separated by a finite energy gap and can be neglected. The procedure is similar also with a defect, but not all the states will enter into the perturbation theory, because the defect will induce an energy gap between the kink states. Namely, the states which have a ferromagnetic bond between the sites $j=N$ and $j=1$ have a different energy from the others.  At $\phi=0$, the states $\ket{N}$ and $\Pi^z\ket{N}$ have the energy
\begin{equation}
E_0=-(N-2)+\zeta,
\end{equation}
while the other kink states have the energy
\begin{equation}
E_0=-(N-2)-\zeta.
\end{equation}

Thus, for $\zeta<0$ the ground states at $\phi=0$ are only $\ket{N}$ and $\Pi^z\ket{N}$. The other kink states are separated by a gap $2\zeta$ and can be neglected, so the perturbation theory is very simple. 
In fact, the perturbation, proportional to $\sum_j\sigma_{j}^y\sigma_{j+1}^y$, does not mix different $\Pi^x$ sectors and is already diagonal in the basis $\ket{N},\Pi^z\ket{N}$.

We conclude that for $\zeta<0$ and small $\phi$ the ground states are (approximately) the states $\ket{N}$ and $\Pi^z\ket{N}$, with magnetization
\begin{equation}
\bra{N}\sigma_j^x\ket{N}=(-1)^{j+1}
\end{equation}
and the one with all spins reversed, respectively. This result explains well the magnetization at Fig.~\ref{fig:1} in the bulk of the system, far from the defects.

For $\zeta>0$ the ground state manifold at $\phi=0$ is \mbox{$2(N-1)$-fold} degenerate. It consists of the states $\ket{j}$ and $\Pi^z\ket{j}$ for $j=1,2,\ldots N-1$. Turning on the perturbation the degeneracy splits. To get the new ground states and the corresponding energies we diagonalize the perturbation in the aforementioned states.

Since the perturbation, proportional to $\sum_j\sigma_{j}^y\sigma_{j+1}^y$, does not mix different $\Pi^x$ sectors we can focus on just the states $\ket{j}$, for $j=1,2,\ldots N-1$. If we include also the state $\ket{N}$, the perturbation is an $N\times N$ cyclic matrix with the elements
\begin{equation}\label{step cyclic}
\bra{k}\sum_{j=1}^N\sigma_j^y\sigma_{j+1}^y\ket{l}= \delta_ {(l-k+2) \ \mathrm{mod}\ N,0}+\delta_ {(l-k-2) \ \mathrm{mod}\ N,0}.
\end{equation}
Without the state $\ket{N}$ the perturbation is a matrix obtained by removing the last row and the last column of the cyclic matrix. It reads
\begin{equation}\label{matrix}
\sum_{j=1}^N\sigma_j^y\sigma_{j+1}^y=
\begin{pmatrix}
0 & 0 & 1 & 0 & \ldots & 0 & 0 & 1 \\
0 & 0 & 0 & 1 & \ldots & 0 & 0 & 0 \\
1 & 0 & 0 & 0 & \ldots & 0 & 0 & 0 \\
0 & 1 & 0 & 0 & \ldots & 0 & 0 & 0\\
\vdots & \vdots & \vdots &  \vdots &  & \vdots & \vdots& \vdots \\
1 & 0 & 0 & 0 & \ldots & 1 & 0 & 0
\end{pmatrix}.
\end{equation}

A similar matrix, obtained by deleting the last row and the last column of the $N\times N$ cyclic matrix with the elements
\begin{equation}
\delta_ {(l-k+1) \ \mathrm{mod}\ N,0}+\delta_ {(l-k-1) \ \mathrm{mod}\ N,0}
\end{equation}
instead of eq.~\eqref{step cyclic} was diagonalized analytically (as a special case) in \cite{Campostrini2015_2}, by writing a recursion relation in $N$ for the characteristic polynomial. We diagonalize the matrix in~\eqref{matrix} in a less dignified way. Based on the similarity with the aforementioned matrix of \cite{Campostrini2015_2} we simply guess the eigenstates. As is easy to check, the normalized eigenstates of the matrix in eq.~\eqref{matrix} are 
\begin{equation}\label{eigenstate 1}
\ket{a_s}=\sqrt{\frac{2}{N}} \sum_{j=1}^{N-1} (-1)^{sj}\sin\Big(\frac{s\pi}{N}j\Big)\ket{j},
\end{equation}
with the eigenvalues $a_s=2\cos\big(\frac{2\pi s}{N}\big)$, and 
\begin{equation}\label{eigenstate 2}
\!\ket{b_s}\!=\!
\begin{cases}
\!\sqrt{\frac{2}{N}} \sum\limits_{j=1}^{N-1} (-1)^{sj+\lfloor \frac{j-1}{2} \rfloor }\sin\left(\frac{s\pi}{N}j\right)\!\ket{j} &\!\!\! N\textrm{mod}4\!=\!1\\
\!\sqrt{\frac{2}{N}} \sum\limits_{j=1}^{N-1} (-1)^{sj+\lfloor \frac{j}{2} \rfloor }\sin\left(\frac{s\pi}{N}j\right)\!\ket{j}  & \!\!\! N\textrm{mod}4\!=\!3
\end{cases}
\end{equation}
with the eigenvalues $b_s=-2\cos\big(\frac{2\pi s}{N}\big)$. 
Here $s$ is one of the possible values from the set $\big\{1,2,\ldots (N-1)/2\big\}$. It follows that the energies associated to the eigenstates in eq.~\eqref{eigenstate 1} and \eqref{eigenstate 2} are respectively
\begin{equation}
\begin{split}
& E_{a,s}=-(N-2)\cos\phi-\zeta +2\sin\phi\cos\Big(\frac{2\pi s}{N}\Big), \\
& E_{b,s}=-(N-2)\cos\phi-\zeta -2\sin\phi\cos\Big(\frac{2\pi s}{N}\Big).
\end{split}
\end{equation}
The parity of the states in eq.~\eqref{eigenstate 1} and \eqref{eigenstate 2} is equal to $\Pi^x=(-1)^{(N-1)/2}$. The states of the opposite parity are constructed, of course, by applying the $\Pi^z$ operator.

Thus, the $2N$-fold degenerate ground state manifold splits, for small $\phi$, into a band of states, with a two-fold degenerate ground state manifold and an energy gap between the states that closes as $1/N^2$. For $\phi>0$ the ground states are $\ket{g}=\ket{a_{s}}$ for $s=(N-1)/2$ and $\Pi^z\ket{g}$, while for $\phi<0$ the ground states are $\ket{g}=\ket{b_s}$ for $s=(N-1)/2$ and $\Pi^z\ket{g}$. After a bit of straightforward algebra, using
\begin{equation}
\bra{l}\sigma_j^x\ket{l}=\begin{cases}
(-1)^{l+j+1}, \quad & l=1,2\ldots,j-1\\
(-1)^{l+j}, \quad & l=j,j+1,\ldots N
\end{cases},
\end{equation}
we find that the magnetization in the ground state $\ket{g}$ is, for both signs of $\phi$,
\begin{equation}\label{magnetization perturbation finite}
\bra{g}\sigma_j^x\ket{g}=\frac{(-1)^j\sin\big[\frac{\pi}{N}\big(j-\frac{1}{2}\big)\big]}{N\sin\big(\frac{\pi}{2N}\big)}+\frac{1}{N}.
\end{equation}
In the ground state $\Pi^z\ket{g}$ the magnetization acquires, of course, an additional minus sign. The obtained order is in agreement with the numerical results on the magnetization in the presence of an antiferromagnetic defect, presented in Fig.~\ref{fig:1}. Note that, for large $N$, the magnetization in eq.~\eqref{magnetization perturbation finite} approximates
\begin{equation}\label{magnetization incommensurate}
\braket{\sigma_j^x}_g=(-1)^j\frac{2}{\pi}\sin\Big(\frac{\pi}{N}j\Big),
\end{equation}
which is the incommensurate AFM order present for $\phi\in(0,\pi/4)$ in the absence of the defect \cite{Maric20_2}. The magnetization is modulated in such a way to achieve zero value where the defect is placed.

\subsection{Perturbation theory around $\zeta=0$}\label{sec perturbation 2}

In this section by using perturbation theory around $\zeta=0$ we find which of the four-fold degenerate ground states of the region $\phi\in(0,\pi/4)$ are selected in the limit of a small antiferromagnetic defect $\zeta\to 0^+$. For this task we treat the term $\zeta\sigma_N^x\sigma_1^x$ in eq.~\eqref{Hamiltonian for perturbation theory} as a perturbation. The model with $\zeta=0$ has been solved in details in \cite{Maric20_2} and we use the same notation. Thus, while before we used the kink states as basis for the perturbation, here we employ the four ground states states determined in \cite{Maric20_2}.

For $\zeta=0$ the ground state manifold is spanned by states $\ket{p},\ket{-p},\Pi^x\ket{p},\Pi^x\ket{-p}$ which are simultaneous eigenstates of the Hamiltonian in eq.~\eqref{Hamiltonian for perturbation theory}, with $\zeta=0$, the parity operator $\Pi^z$ (with eigenvalues, respectively $\Pi^z=-1,-1,1,1$) and the translation operator $T$ (with eigenvalues, respectively $T=e^{\imath p},e^{-\imath p},e^{\imath p},e^{-\imath p}$). Here $p=\pi/2+(-1)^{(N+1)/2}\pi/2N$ is the momentum of the states. Above the ground states there is a band of states, with the energy gap closing as $1/N^2$. 

To find which ground state vectors are selected in the limit of a small defect we diagonalize the perturbation $\zeta\sigma_N^x\sigma_1^x$ in the basis of the four ground states above. We are going to neglect all the excited states of the model, including those belonging to the lowest-energy band. This is not justified in general, because of the gapless nature of the system, but the procedure is going to yield the results in agreement with numerics, as we comment in the end. Since the perturbation does not mix different $\Pi^x$ sectors it is sufficient to focus on the subspace spanned by $\ket{p},\ket{-p}$. Thus, we need to compute and diagonalize the matrix
\begin{equation}
\sigma_N^x\sigma_1^x=
\begin{pmatrix}
\bra{p}\sigma_N^x\sigma_1^x\ket{p} & \bra{p}\sigma_N^x\sigma_1^x\ket{-p}\\
\bra{-p}\sigma_N^x\sigma_1^x\ket{p} & \bra{-p}\sigma_N^x\sigma_1^x\ket{-p}
\end{pmatrix}.
\end{equation}

The elements of the perturbation matrix are computed using the Majorana fermions representation of the spin operators, in terms of which
\begin{equation}\label{perturbation Majorana}
\sigma_N^x\sigma_1^x=\Pi^z(-\imath A_1B_N),
\end{equation}
and using the representation of the Majorana fermions in terms of Bogoliubov fermions $a_q$, that can be obtained from the exact solution presented in \cite{Maric20_2}. We have
\begin{equation}\label{Majorana Bogoliubov}
\begin{split}
& A_j=\frac{1}{\sqrt{N}}\sum\limits_{q\in\Gamma^-}(a_q^\dagger+a_{-q})e^{\imath\theta_q}e^{-\imath q j},\\
& B_j=\frac{1}{\sqrt{N}}\sum\limits_{q\in\Gamma^-} \imath (a_q^\dagger-a_{-q})e^{-\imath\theta_q}e^{-\imath q j},
\end{split}
\end{equation}
where $\Gamma^-=\{2\pi k/N : k=0,1,\ldots N-1\}$ and the Bogoliubov angle $\theta_q$ is defined as
\begin{equation}\label{arctan}
\!\!\theta_{q}\!=\!\tan^{-1}\frac{|\sin \phi\! +\! \cos \phi \ e^{i2q}| \!-\!(\sin\phi\!+\!\cos\phi)\cos q}{(-\sin\phi \!+\! \cos\phi)\sin q}
\end{equation}
for $q\neq 0$ and $\theta_0=0$.
In terms of Bogoliubov fermions the ground states are given by $\ket{\pm p}=a_{\pm p}^\dagger\ket{0^-}$, where $\ket{0^-}$ is the vacuum state, satisfying $a_q \ket{0^-}=0,\ q\in\Gamma^-$. Using eq.~\eqref{perturbation Majorana}, \eqref{Majorana Bogoliubov} and this ground states representation we get the matrix elements of the perturbation. 
For $\bra{p}\sigma_N^x\sigma_1^x\ket{p}=\bra{-p}\sigma_N^x\sigma_1^x\ket{-p}$ we recover
\begin{equation}
\bra{p}\sigma_N^x\sigma_1^x\ket{p}=\frac{2}{N}\cos(2\theta_p-p)-\frac{1}{N}\sum\limits_{q\in\Gamma^-}e^{\imath (2\theta_q-q)}
\end{equation}
while for $\bra{p}\sigma_N^x\sigma_1^x\ket{-p}$
\begin{equation}
 \bra{p}\sigma_N^x\sigma_1^x\ket{-p}=\frac{2}{N}e^{-\imath p}.
\end{equation}
Then, diagonalizing the perturbation matrix we obtain the eigenstates
\begin{equation}\label{xipm}
\ket{\xi_\pm}=\frac{1}{\sqrt{2}}(\ket{p}\pm e^{\imath p}\ket{-p}),
\end{equation}
which are also even/odd under the mirror symmetry crossing the site $\frac{N+1}{2}$ (see the Supplementary Information of~\cite{Maric20_2}). These states have energies
\begin{equation}
\!\!\!E_\pm\!=\!E_0\!-\!\zeta\frac{1}{N}\!\!\sum\limits_{q\in\Gamma^-}\!\!e^{\imath (2 \theta_q-q)}+\zeta\frac{2}{N}\cos(2\theta_p\!-\!p)\pm\zeta\frac{2}{N},
\end{equation}
where $E_0$ is the ground state energy of the unperturbed model. For the antiferromagnetic defect $\zeta>0$ the state $\ket{g^-}\equiv\ket{\xi_-}$ is lower in energy and, therefore, (approximately) the new ground state, belonging to $\Pi^z=-1$ sector. Of course, the ground state belonging to $\Pi^z=+1$ sector is $\ket{g^+}=\Pi^x\ket{\xi_-}$.

The magnetization can be computed using the same techniques as in \cite{Maric20_2}, that employ the translation operator. Denoting $\ket{g}=\frac{1}{\sqrt{2}}(1+\Pi^x)\ket{g^-}$, we get
\begin{eqnarray}\label{magnetization spatial}
\braket{\sigma_j^x}_g&=&\bra{g^-}\sigma_j^x\Pi^x\ket{g^-}\nonumber \\
&=&(-1)^{j}(-1)^{\frac{N-1}{2}}\sin\Big[\frac{\pi}{N}\Big(j-\frac{1}{2}\Big)\Big]\bra{p}\sigma_N^x\Pi^x\ket{-p}\nonumber\\
& & +\bra{p}\sigma_N^x\Pi^x\ket{p},
\end{eqnarray}
where we have used that $\bra{p}\sigma_j^x\Pi^x\ket{-p} = e^{- \imath 2 p j} \bra{p}\sigma_N^x\Pi^x\ket{-p}$ and that $\bra{p}\sigma_N^x\Pi^x\ket{-p} = \bra{-p}\sigma_N^x\Pi^x\ket{p}$ from~\cite{Maric20_2}.
The matrix elements encountered in this expression have also been computed in \cite{Maric20_2}. It has been found numerically that in the thermodynamic limit $N\to\infty$ we have
\begin{eqnarray}
\bra{p}\sigma_N^x\Pi^x\ket{-p} & = & \frac{2}{\pi}(1-\tan^2\phi)^{1/4},
\label{matrixelements1} \\
\bra{p}\sigma_N^x\Pi^x\ket{p} & = & 0,
\label{matrixelements2}
\end{eqnarray}
which gives the magnetization
\begin{equation}\label{thermomagnetization}
\!\!\!\braket{\sigma_j^x}_g\!=\!\frac{2 (-1)^{\frac{N-1}{2}+j}}{\pi}\!(1-\tan^2\!\phi)^{\frac{1}{4}}\sin\left[\frac{\pi}{N}\left(j\!-\!\frac{1}{2}\right)\right]
\end{equation}
The obtained magnetization generalizes eq.~\eqref{magnetization incommensurate} to the whole region $\phi\in(0,\pi/4)$ (the factor $(-1)^{(N-1)/2}$ of difference arises because of the different parities of the involved states) and describes well our numerical results. Note that, since the states in eq.~\eqref{xipm} are eigenstates of the mirror symmetry across the site $\frac{N+1}{2}$, the magnetization pattern they generate must be even under such transformation, a property present in that is in eq.~\eqref{thermomagnetization}, but not in eq.~\eqref{magnetization incommensurate}.

Since we have performed two different perturbation theories, we can check their agreement in the regime where both applies and at finite sizes.
From \cite{Maric20_2} we know that in the limit $\phi\to 0^+$ we have
\begin{align}
&\bra{p}\sigma_N^x\Pi^x\ket{-p}=\frac{1}{N\sin\big(\frac{\pi}{2N}\big)},\\
&\bra{p}\sigma_N^x\Pi^x\ket{p}=(-1)^{\frac{N-1}{2}}\frac{1}{N}.
\end{align}
Sticking this into eq.~\eqref{magnetization spatial} gives us exactly eq.~\eqref{magnetization perturbation finite}, up to factor $(-1)^{(N-1)/2}$ that arises from the parities of the involved states. We can thus infer that eq. (\ref{matrixelements1},\ref{matrixelements2}) are similarly corrected at finite sizes and thus arrive at eq.~\eqref{strange_mag} in the main text.

The perturbation theory done in this section describes well our numerical results in the region $\phi\in(0,\pi/4)$ in the case of an antiferromagnetic defect and shows to which ground states out of the four-fold degenerate manifold the discovered order corresponds. The same perturbation theory would not be successful in describing the order in all cases of $\phi$ and $\zeta$. The reason is that due to the gapless nature of the system it is not justified to neglect the low-lying states of the model in the perturbation theory, so the procedure does not have to give the right results in general.

\end{document}